\documentstyle[twocolumn,aps,prc,psfig]{revtex}% Use to check format & length.
\begin{document}

\title{Light cluster production in E/A = 61 MeV $^{36}$Ar + $^{112,124}$Sn reactions}

\author{
R.~Ghetti$^{1}$, 
J.~Helgesson \\
{\small \it School of Technology and Society, Malm\"o University, S-205 06 Malm\"o, Sweden}\\
V.~Avdeichikov, 
B.~Jakobsson \\
{\small \it Department of Physics, Lund University, Box 118, S-22100 Lund, Sweden}\\
N.~Colonna, 
G.~Tagliente \\
{\small \it INFN and Dip.\ di Fisica, V.~Amendola 173, I-70126 Bari, Italy}\\
H.W.~Wilschut,
V.L.~Kravchuk$^{2}$\\ 
{\small \it Kernfysisch Versneller Instituut, Zernikelaan 25 NL 9747 AA Groningen, The Netherlands}\\
}
\maketitle

\begin{abstract}
Experimental kinetic energy distributions  
and small-angle two-particle correlation functions 
involving deuterons and tritons are compared 
for $^{36}$Ar+ $^{112,124}$Sn collisions at E/A = 61 MeV 
(i.e.\ for systems similar in size, but with different isospin content). 
A larger triton yield is observed from the more neutron-rich system, 
as predicted by IBUU simulations, 
while the emission times of the light clusters are found 
to be the same for the two Sn-target systems. 
For both systems, the time sequence 
$\,\tau_{d}\,$$<$$\,\tau_{p}\,$$<$$\,\tau_{t}$,
is deduced for charged particles emitted from the 
intermediate velocity source. 
\end{abstract}

\vspace{0.1cm}
\noindent
PACS number(s): 25.70.Pq

\vspace{0.1cm}
{\small 
\noindent
$^1$Corresponding Author: Roberta Ghetti, 
School of Technology and Society, Malm\"o University, Sweden}
{\it E-mail address}: roberta.ghetti@ts.mah.se\\
$^{2}$Present address: LNL, Legnaro, Italy.
\normalsize
%============================================================
\section{Introduction}
%============================================================
Among the challenges in nuclear physics today,
is to understand the nuclear interaction for systems
with an exotic composition of neutrons and protons
(which may occur in the $r$-process of nucleosynthesis 
and in neutron stars). 
One of the topics in this challenge is to understand
the isospin dependence of the nuclear equation of state (EOS), 
and in particular its density dependence
\cite{Nova,Mueller,Colonna,LiPRL00,Ma,Liu}. 
In this difficult task many observables are needed
to put as many constraints as possible on the nuclear interaction.
Among several suggested observables,
\cite{LiPRL97,Baran02,LiKo97,Xu,Tan,Tsang,LiPRC01}, 
the two-nucleon correlation function has been 
considered in Refs.\ \cite{Chen,Chen2}. 
 
Experimentally, isospin effects in 
two-particle correlation functions were investigated by our group 
in Ref.\ \cite{Rapid}. The experimental data were from 
E/A~=~61~MeV $^{36}$Ar-induced semi-peripheral 
collisions on isotope-separated $^{112,124}$Sn targets. 
Stronger $np$ correlations were 
found for the Ar + $^{124}$Sn system, pointing 
to a shorter neutron emission time from the more neutron-rich 
system. Smaller isospin effects were seen also in $pp$, 
$nd$ and $nt$ correlation functions \cite{Rapid}.
In addition, a study of the emission time sequence of neutrons, protons, and 
deuterons  
from the same reactions was carried out by our group 
in Ref.\ \cite{Led04}, utilizing the method of 
the velocity-gated correlation functions. That investigation 
revealed a sensitivity of the particle emission time sequence 
to the isospin content of the emitting source, particularly for 
intermediate velocity source emission and emission from the target residues. 

In this paper we present complementary information 
to that of Refs.\ \cite{Rapid,Led04}, from the 
same experimental data set, focusing on deuterons and tritons.
Indeed, also light cluster production is expected to be a sensitive probe 
for the isospin dependence of the EOS. 
In the theoretical investigation of Ref.\ \cite{Chen-LCP} 
(where IBUU was coupled with a coalescence model for cluster production), 
it was found that the nuclear symmetry energy 
significantly affects the production of light clusters 
in heavy ion collisions. 
More deuterons and tritons are produced 
with a stiff nuclear symmetry energy than with a soft. 
This is because the stiff symmetry energy 
induces a stronger pressure in the reaction system, 
and thus causes an earlier emission of neutrons and protons 
leading to stronger correlations among nucleons, 
and, consequently, larger yields of deuterons and tritons. 

%============================================================
\section{Experimental details}
%============================================================
The experiment was performed at KVI (Groningen).  
The interferometer consisted of 16 CsI(Tl)  
detectors for light charged particles, mounted in the angular 
range 30$^{\rm o}$ $\le$ $\theta$ $\le$ 114$^{\rm o}$, 
and 32 liquid scintillator neutron detectors, 
mounted behind the ``holes'' 
of the CsI array, in matching positions  
to provide the $np$ interferometer \cite{NimTof}. 
Energy thresholds for protons, deuterons, and tritons 
in the CsI(Tl) detectors 
were 8, 11, and 14~MeV respectively.
Details about the experimental setup and the particle energy 
determination are given in Refs.\ \cite{NimTof,NiAl,Volly}. 

From our previous analyses \cite{Rapid,Led04}, including the 
source analysis of Ref.\ \cite{Volly}, we have confirmed that 
the emission of light particles from 61 A MeV 
(semi-peripheral) heavy ion reactions 
originates from (at least) three sources: projectile and target residue 
evaporative sources (statistical evaporation) 
and intermediate velocity source. The intermediate source 
represents dynamical emission, which is described either by early 
nucleon-nucleon collisions or by other pre-equilibrium processes  
\cite{Montoya,Toke95,Laro97,Luka97,Pawl98,Laro99,Plag99,Mila00,InAl03}. 
Within the multi-source reaction mechanism described above, 
composite particles, like deuterons and tritons, are believed to be 
predominantly emitted from the dynamical emission source \cite{Lanzano,Lefort}, 
where they are formed by a coalescence mechanism \cite{Volly,Coalescence}. 

%============================================================
\section{Results}
\begin{figure}
\centerline{\psfig{file=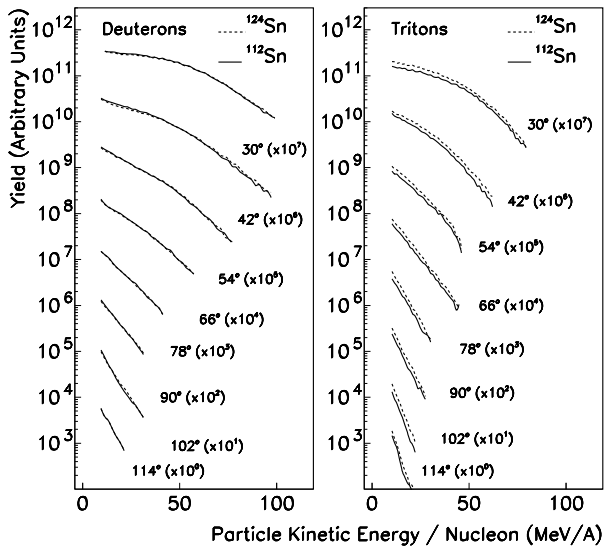,height=7.5cm,angle=0}}
\caption{\small
Comparison of the $d$ and $t$ kinetic energy 
yields measured in $^{36}$Ar+$^{124}$Sn (dashed lines) 
and $^{36}$Ar+$^{112}$Sn (full lines) at the 
angles indicated in the figure.
The yields are arbitrarily shifted in the y-axis 
as indicated in the figure.
}
\label{yields}
\end{figure}

%============================================================
Figure \ref{yields} compares the kinetic energy distributions of deuterons 
and tritons measured in $^{36}$Ar+$^{112}$Sn and $^{36}$Ar+$^{124}$Sn 
collisions. 
One can notice that the yields of the deuteron spectra are the
same for the two Sn-targets, while the triton yield from the 
more neutron-rich  $^{36}$Ar+$^{124}$Sn system is 
enhanced (by a factor 1.1 to 1.5)
over the whole angular and energy range considered. 

To further investigate the deuteron and triton production, we 
present the two-particle correlation functions 
involving deuterons and tritons. 
The correlation function, 
$C(\vec{q},\vec{P}_{tot})$ 
= 
$k N_c(\vec{q},\vec{P}_{tot}) $ 
/ 
$N_{nc}(\vec{q},\vec{P}_{tot})$,
is generated by dividing the coincidence yield ($N_c$) 
by the yield of non-correlated events ($N_{nc}$) 
constructed from the product of the 
singles distributions \cite{NPA00}. 
The relative momentum is, 
$\vec{q} = \mu (\vec{p}_{1} / m_1 - \vec{p}_{2} / m_2)$, 
where $\mu$ is the reduced mass, 
and the total momentum is,
$\vec{P}_{tot} = \vec{p}_{1}+\vec{p}_{2}$.
The correlation function is normalized to unity at large values of $q$, 
(200 $<$ $q$ $<$ 300 MeV/c for all particle pairs) 
where no correlations are expected.

The $pp$, $np$, and $pd$ correlation functions 
have already been presented in Refs.\ \cite{Rapid,Led04}. 
In those analyses, particles emitted by the intermediate velocity source 
were enhanced by selecting high- and intermediate-$P_{tot}$ pairs, 
while particles emitted by the target residues were enhanced by selecting 
low-$P_{tot}$ pairs in the target frame. 
Further selection of different sources 
was obtained by applying angular gates. 
The $pp$ and $np$ correlation functions 
were found to be very sensitive to the applied gates,
indicating that both neutrons and protons are
emitted from several sources.
In contrast, deuterons and tritons mainly originate 
from common sources with a small spread in source velocity
($v_{source} \approx$ 0.18 c) \cite{Volly}.
In agreement with those findings, 
we find, in the present work,  
that the $dd$, $tt$, and $dt$ correlation functions 
are quite insensitive to applying total-momentum and angular gates. 
We have therefore chosen to present angle and total-momentum-integrated 
$dd$, $tt$ and $dt$ correlation functions in this paper.

\begin{figure}
\centerline{\psfig{file=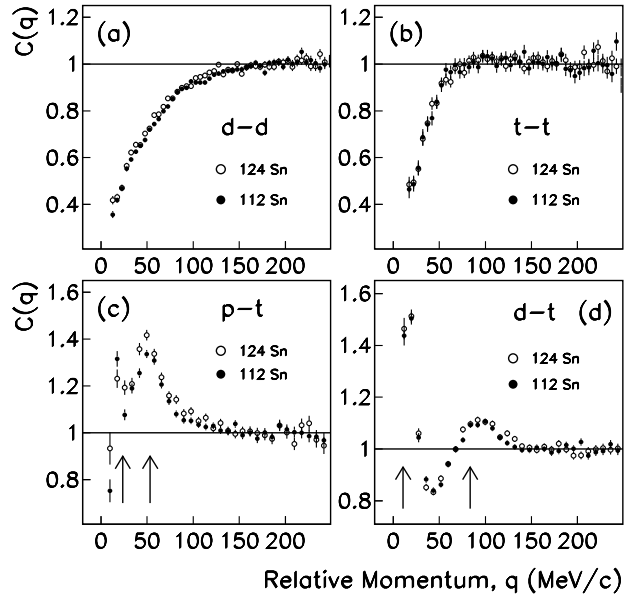,height=8.cm,angle=0}}
\caption{\small
From $^{36}$Ar+$^{124}$Sn (open circles) 
and $^{36}$Ar+$^{112}$Sn (filled circles), 
$dd$ (a), $tt$ (b), and $dt$ (d) 
angle- and total-momentum-integrated correlation functions.
A high-total-momentum gate is applied to the $pt$ (c) 
correlation function. 
}
\label{cfs}
\end{figure}

Figure \ref{cfs}(a,b) compares the $dd$ (a) and $tt$ (b) 
correlation functions measured from the two Sn-target systems. 
One can see that the Coulomb hole dominates both correlations at 
low values of the relative momentum. 
For both $dd$ and $tt$, the width of the Coulomb hole 
is the same for the two Sn-targets.
This feature indicates that the average emission time of deuterons 
and tritons is the same for the two systems. 
The width of the Coulomb hole is larger for
$dd$ than for $tt$, which is consistent
with a larger average emission time for
tritons than for deuterons.

The $dt$ correlation function in Fig.\ \ref{cfs}d 
is very similar for the two Sn-targets.
It is governed by
strong resonances at 
$q$ $\approx$ 10.8 MeV/c 
and
$q$ $\approx$ 83.5 MeV/c, 
due to the excited states, 
$E^*$ = 16.75 MeV ($\Gamma_{cm}$ = 0.076 MeV) 
and
$E^*$ = 19.8 MeV ($\Gamma_{cm}$ = 2.5 MeV)
of $^5$He \cite{Tilley1}.

We now turn to the correlation functions involving also protons.
The $pd$ correlation function, that was 
presented in Refs.\ \cite{Rapid,Led04}, 
is characterized by a pronounced anti-correlation at small relative 
momenta, due to final state Coulomb interaction. 
The anti-correlation is more pronounced for 
particle pairs selected 
by high- and intermediate-total-momentum gates, 
than for pairs selected by low-total-momentum gate.

The $pt$ correlation function, shown in Fig.\ \ref{cfs}c, 
contains contributions of excited states of $^4$He at 
$q$ $\approx$ 23.6 MeV/c ($E^*$ = 20.21 MeV, $\Gamma_{cm}$ = 0.5 MeV),
$q$ $\approx$ 41.0 MeV/c ($E^*$ = 21.01 MeV, $\Gamma_{cm}$ = 0.84 MeV),
and
$q$ $\approx$ 53.4 MeV/c ($E^*$ = 21.84 MeV, $\Gamma_{cm}$ = 2.01 MeV) 
\cite{Tilley},
while higher excited states of $^4$He are much broader and
not visible in Fig.\ \ref{cfs}c.
The positions of the 1$^{st}$ and 3$^{rd}$ resonances 
are marked by arrows. 

In contrast with $dd$, $tt$, and $dt$, the $pd$ and $pt$ 
correlation functions  
are sensitive to applying angle- and total-momentum gates. 
This is because protons are emitted from several sources \cite{Volly}.
In order to consistently investigate emission 
from the intermediate velocity source, 
we have therefore applied a high-total-momentum gate 
to the $pt$ correlation function (shown in Fig.\ \ref{cfs}c). 
This selection generates a correlation function characterized 
by stronger resonance peaks as compared to those obtained 
by lower total momentum  gates. 
Inspection of Fig.\ \ref{cfs}c reveals also 
a small difference 
in the correlation functions for the two Sn isotopes. 
The more neutron-rich system leads to a slightly 
stronger resonance peak 
for the higher excited states of $^4$He, 
while the opposite behavior is seen 
for the lowest excited state.

Further information about the emission times and chronologies
can be obtained by means of particle-velocity-gated correlation functions 
of non-identical particles \cite{Lednicky,PRL-01,Gourio}. 
This adds an important and novel piece of information 
to the picture of the reaction mechanism emerging from 
previous studies \cite{Rapid,Led04}.
For non-identical particles, $a$ and $b$, we construct 
the correlation functions $C_a(q)$, gated on pairs 
with $v_a > v_b$, and $C_b(q)$, gated on pairs 
with $v_b > v_a$. 
The particle velocities are calculated 
in the frame of the emitting source, 
and the same normalization constant,
calculated from the ungated correlation function, 
is utilized for both $C_a$ and $C_b$ \cite{PRL-01}. 
If two particles $a$ and $b$ are emitted independently from a source,
with $a$ later (earlier) than  $b$, 
then the ratio $C_a/C_b$ will show a peak (dip) 
in the region of $q$ where there is a correlation, 
and a dip (peak) in the region of $q$ where there is an 
anti-correlation. 

To correctly interpret the experimental results, it is
important to realize that
the resonance peaks may a priori have two different origins 
\cite{Pochodzalla,Deyoung}:
\begin{enumerate}
\item
  From processes where an unstable fragment formed in the reaction 
  decays into two particles which are measured in coincidence 
  (e.g.\ $^5$He $\rightarrow$ $d$ + $t$).
  The energy and momentum conservation in the decay
  determines the location of the resonance peak.

\item
  From final state interactions between particles
  emitted independently from a source 
  (e.g.\ intermediate velocity source, 
   target and projectile residues). 

\end{enumerate}
If the two particles are emitted independently, the velocity-gated 
correlation functions contain information on the time sequence, 
but this is not the case if the two particles originate from the 
two-body decay of an unstable fragment emitted in the reaction. 
In the latter case, 
the two particle velocities are determined by momentum conservation, 
and in the rest system of the decaying fragment, 
the lightest particle will always get the highest velocity. 
In this case, the velocity-gated correlation function 
obtained with the condition that the lightest particle 
has the largest velocity, 
should exhibit the strongest correlation or anti-correlation
   (see Ref.\ \cite{Interpr-vel-gate} for further details). 
When we in some cases observe the opposite behavior 
(namely that the gate, 
 where the {\it heaviest} particle has the largest velocity,
 leads to a stronger correlation or anti-correlation) 
we can reliably conclude that this behavior is dominated by 
a mechanism other than two-body decay. We attribute this 
effect to the interaction of independently emitted particles, 
and in this case, we use the velocity-gated correlation 
function to obtain information on the time sequence of the 
independently emitted particles.
\begin{figure}
\centerline{\psfig{file=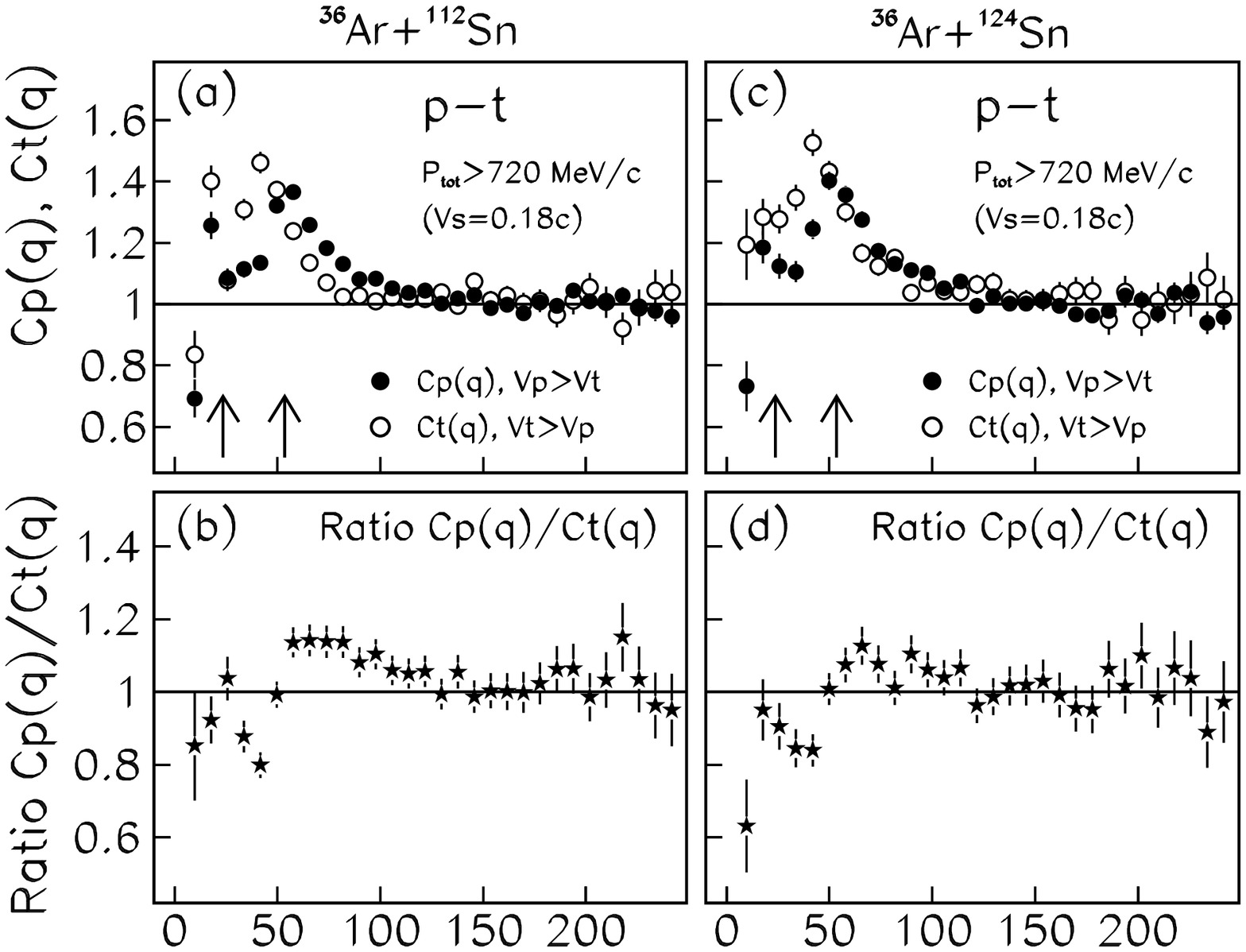,height=7.0cm,angle=0}}
\vspace{-1.2cm}
\centerline{\psfig{file=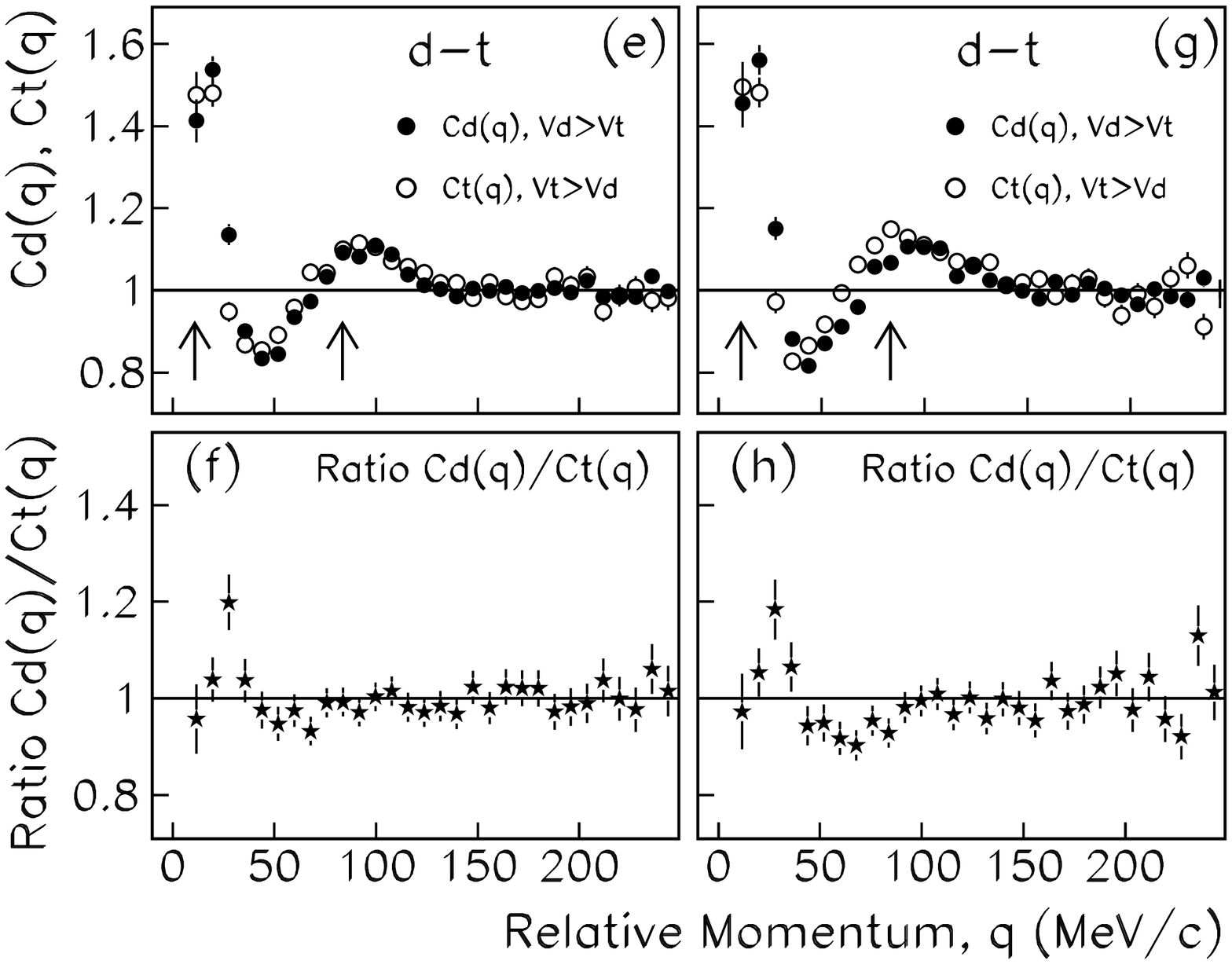,height=7.0cm,angle=0}}
\caption{
From $E/A$ = 61 MeV  
$^{36}$Ar + $^{112}$Sn (left column) and  $^{124}$Sn 
(right column) collisions, 
particle-velocity-gated (filled and open circles) 
$pt$ (a,c) and $dt$ (e,g) correlation functions, 
and their ratios (b,d,f,h).  
}
\label{led}
\end{figure}

The study of the particle-velocity-gated $pd$ correlation function 
can be found in Ref.\ \cite{Led04}. 
The extracted emission time sequence is that deuterons are, 
on average, emitted earlier than protons for both Sn-target systems
(consistent with the findings 
 of previous experimental investigations \cite{PRL-01,Gourio}).
Note that this result 
refers to the average emission time, and it does not exclude 
that prompt protons are emitted before any deuteron is emitted. 

The particle-velocity-gated $p$$t$ correlation function is  
shown in Fig.\ \ref{led}a-d. 
In the momentum region below $q$ $\approx$ 50 MeV/c,
the enhancement of 
$C_p$ ($v_p > v_t$) over $C_t$ ($v_t > v_p$), 
expected from momentum conservation of the two-body 
decay of $^4$He, is not observed. 
This indicates that the behavior of the $p$$t$ correlation function 
in this region is dominated by independently emitted pairs. 
The emission chronology that can be deduced from this region 
is that protons are emitted earlier than tritons. 
Inspection of the  $C_p/C_t$ ratios (Fig.\ \ref{led}b,d) 
reveals that the deduced $p$$t$ emission chronology 
is qualitatively similar for the two Sn-target systems.  

The particle-velocity-gated $dt$ 
correlation function in Fig.\ \ref{led}e-h,
shows an enhancement of the $C_d$ correlation function
that is expected from the two-body decays of $^5$He. 
The behavior of the  $C_d/C_t$ ratio 
(Fig.\ \ref{led}f,h) is therefore of delicate interpretation, 
since it may be dominated by the two-body resonance decays of $^5$He,
and it does not led itself  
to deduce the emission chronology in this case. 

%============================================================
\section{Summary}
%============================================================
Kinetic energy spectra and two-particle 
correlation functions of deuterons and tritons 
have been investigated in 
E/A = 61 MeV $^{36}$Ar+$^{112,124}$Sn reactions.
The yield of tritons is larger from the more neutron-rich system, 
while no differences are found in the deuteron yield. 
The $dd$, $tt$ and $dt$ correlation functions are rather insensitive 
to applying angle- and total-momentum gates,
supporting the interpretation \cite{Volly} 
that the origin of emission of deuterons and tritons 
is a coalescence mechanism taking place 
in the intermediate velocity source created in the early dynamical phase.
For $pd$ and $pt$ pairs, instead, the 
emission from the intermediate velocity source can be 
enhanced by selecting high-total-momentum pairs. 

From Ref.\ \cite{Led04} and the analysis 
of the velocity-gated $pt$ correlation function presented here, 
we find that, for the intermediate velocity source, 
deuterons are on average emitted before protons, which
in turn are emitted before tritons,
$\,\tau_{d}\,$$<$$\,\tau_{p}\,$$<$$\,\tau_{t}$. 
This is also consistent with the behavior of the
$d$$d$ and $t$$t$ correlation functions.

The isospin effects in deuteron and triton emission, 
appear to be weak for the studied systems. 
Apart for a larger triton yield from the more neutron-rich system, 
the emission chronology appears to be the same, 
as well as the average deuteron and triton emission times. 
One possible interpretation is that the larger yield 
of tritons in the $^{36}$Ar+$^{124}$Sn system
is formed in the same amount of time as in the
$^{36}$Ar+$^{112}$Sn system, because of a larger abundance 
of neutrons emitted early in the $^{36}$Ar+$^{124}$Sn reaction.
\\

{\bf Acknowledgements}\\
The author RG wishes to thank the School of Technology and 
Society of Malm\"o University. 

%------------------------------------------------------------------------------


\begin{thebibliography}{9}
\bibitem{Nova} \it Isospin Physics in Heavy-Ion 
               Collisions at Intermediate Energies, \rm 
               edited by Bao-An Li and W.U.\ Schr\"oder 
               (Nova Science Publishers, Inc., New York, 2001).

\bibitem{Mueller} H.\ M\"uller and B.\ Serot, Phys.\ Rev.\ C {\bf 52}, 2072 (1995).

\bibitem{Colonna}  M.\ Colonna, {\it et al.}, 
                   Phys.\ Rev.\ C {\bf 57}, 1410 (1998).

\bibitem{LiPRL00} Bao-An Li, Phys.\ Rev.\ Lett.\ {\bf 85}, 4221 (2000).

\bibitem{Ma} Y.G.\ Ma, {\it et al.}, Phys.\ Rev.\ C {\bf 60}, 024607 (1999).

\bibitem{Liu} J.-Y.\ Liu {\it et al.}, Phys.\ Rev.\ C {\bf 63}, 054612 (2001).

\bibitem{LiPRL97} Bao-An Li, {\it et al.}, % C.M.\ Ko, Z.Z.\ Ren,
                  Phys.\ Rev.\ Lett.\ {\bf 78}, 1644 (1997).

\bibitem{Baran02}   V.\ Baran, {\it et al.},
                    Nucl.\ Phys.\ {\bf A703}, 603 (2002).

\bibitem{LiKo97} Bao-An Li and C.M.\ Ko, Nucl.\ Phys.\ {\bf A618}, 498 (1997).

\bibitem{Xu} H.S.\ Xu, {\it et al.} Phys.\ Rev.\ Lett.\ {\bf 85}, 716 (2000).

\bibitem{Tan} W.P.\ Tan, {\it et al.}, Phys.\ Rev.\ C {\bf 64}, 051901(R) (2001).

\bibitem{Tsang} M.B.\ Tsang, {\it et al.}, Phys.\ Rev.\ Lett.\ {\bf 86}, 5023 (2001).

\bibitem{LiPRC01} Bao-An Li, {\it et al.}, % A.T.\ Sustich, B.\ Zhang, 
                  Phys.\ Rev.\ C {\bf 64}, 054604 (2001).

\bibitem{Chen}    L.W.\ Chen, {\it et al.}, %V.\ Greco, C.M.\ Ko, Bao-An Li, 
                  Phys.\ Rev.\ Lett.\ {\bf 90}, 162701 (2003).

\bibitem{Chen2}    L.W.\ Chen, {\it et al.}, % V.\ Greco, C.M.\ Ko, Bao-An Li, 
                   Phys.\ Rev.\  C {\bf 68}, 014605 (2003).

\bibitem{Rapid}  R.\ Ghetti, {\it et al.},
                 Phys.\ Rev.\  C {\bf 69}, 031605(R) (2004).

\bibitem{Led04}    R.\ Ghetti, {\it et al.},
                   Phys.\ Rev.\  C {\bf 70}, 034601 (2004).

\bibitem{Chen-LCP} L.W.\ Chen, {\it et al.}, % V.\ Greco, C.M.\ Ko, Bao-An Li, 
                   Phys.\ Rev.\  C {\bf 68}, 017601 (2003).

\bibitem{NimTof}   R.\ Ghetti, {\it et al.}, Nucl.\ Inst.\ Meth.\  {\bf A 516}, 492 (2004). 

\bibitem{NiAl}     R.\ Ghetti, {\it et al.}, Phys.\ Rev.\ Lett.\ {\bf 91}, 092701 (2003).

\bibitem{Volly}    V.\ Avdeichikov, {\it et al.}, %R.\ Ghetti, P.\ Golubev, B.\ Jakobsson, 
                   Nucl.\ Phys.\ {\bf A 736}, 22 (2004).

%  NECK EXP--------------------------------------------------------------
\bibitem{Montoya} C.P.\ Montoya, {\it et al.}, Phys.\ Rev.\ Lett.\ {\bf 73}, 3070 (1994). 

\bibitem{Toke95} J.\ T\~oke, {\it et al.}, Phys.\ Rev.\ Lett.\ {\bf 75}, 2920 (1995). 

\bibitem{Laro97} Y.\ Larochelle, {\it et al.}, Phys.\ Rev.\ C {\bf 55}, 1869 (1997).

\bibitem{Luka97} J.\ {\L}ukasik, {\it et al.}, % (The INDRA Collaboration), 
                 Phys.\ Rev.\ C {\bf 55}, 1906 (1997).

\bibitem{Pawl98} P.\ Pawloswki, {\it et al.}, Phys.\ Rev.\ C {\bf 57}, 1771 (1998).

\bibitem{Laro99} Y.\ Larochelle, {\it et al.}, Phys.\ Rev.\ C {\bf 59}, R565 (1999).

\bibitem{Plag99} E.\ Plagnol, {\it et al.}, %(The INDRA Collaboration), 
                 Phys.\ Rev.\ C {\bf 61}, 014606 (1999).

\bibitem{Mila00} P.M.\ Milazzo, {\it et al.}, Nucl.\ Phys.\ {\bf A 703}, 466 (2002).

\bibitem{InAl03}  J.\ {\L}ukasik, {\it et al.}, %Indra Collaboration, Aladin Collaboration, 
                  Phys.\ Lett.\ B,  {\bf 566}, 76 (2003). 

\bibitem{Lanzano} G.\ Lanzan{\`o}, {\it et al.}, Nucl.\ Phys.\ {\bf A 683}, 566 (2001).

\bibitem{Lefort} T.\ Lefort, {\it et al.}, Nucl.\ Phys.\ {\bf A 662}, 397 (2000).

\bibitem{Coalescence} W.J.\ Llope, {\it et al.}, Phys.\ Rev.\ C {\bf 52}, 2004 (1995).

\bibitem{NPA00}    R.\ Ghetti, {\it et al.}, Nucl.\ Phys.\ {\bf A674}, 277 (2000).

\bibitem{Tilley1} D.~R.\ Tilley, {\it et al.}, Nucl.\ Phys.\ {\bf A 708}, 3 (2002).

\bibitem{Tilley} D.\ R.\ Tilley, {\it et al.}, Nucl.\ Phys.\ {\bf A 541}, 1 (1992).

\bibitem{Lednicky} R.\ Lednicky,  {\it et al.}, Phys.\ Lett.\ B {\bf 373}, 30 (1996). 

\bibitem{PRL-01} R.\ Ghetti, {\it et al.}, Phys.\ Rev.\ Lett. {\bf 87}, 102701-1 (2001).

\bibitem{Gourio} D.\ Gourio, {\it et al.},  Eur.\ Phys.\ J.\  A {\bf 7}, 245 (2000).

\bibitem{Pochodzalla} J.\ Pochodzalla, {\it et al.}, Phys.\ Rev.\ C {\bf 35}, 1695 (1987).

\bibitem{Deyoung} P.~A.~DeYoung, {\it et al.}, Phys.\ Rev.\ C {\bf 39} (1989) 128; 
 Phys.\ Rev.\ C {\bf 41} (1990) R1885, and references therein.

\bibitem{Interpr-vel-gate} J.\ Helgesson, {\it et al.},
                            submitted to Phys.\ Rev.\ C.

\end{thebibliography}
\end{document}